\documentclass[12pt]{article}

\usepackage[margin=2.5cm]{geometry}

\usepackage{graphicx}
\usepackage[T1]{fontenc}
\usepackage[]{times}
\usepackage[latin1]{inputenc}
\usepackage{booktabs}
\usepackage{geometry}
\usepackage{bbm}
\usepackage[sumlimits, intlimits, namelimits]{amsmath}
\usepackage{amssymb}
\usepackage{amsfonts}
\usepackage[dvips]{epsfig}
\usepackage{amsthm}
\usepackage{fancyhdr}
\usepackage[mathscr]{eucal}
\usepackage{braket}
\usepackage[ansi]{umlaute}
\usepackage{float}
\usepackage{sistyle}
\usepackage{subfigure}
\usepackage{url,comment}

\usepackage{dcolumn}
\newcolumntype{d}[1]{D{.}{.}{#1}}

\def\rz{\mathbb{R}}

\def\pz{\mathbb{P}}

\def\bF{\mathbf{F}}
\def\bY{\mathbf{Y}}
\def\by{\mathbf{y}}
\def\bx{\mathbf{x}}
\def\bZ{\mathbf{Z}}
\def\bz{\mathbf{z}}

\def\bC{\mathbf{C}}

\usepackage{natbib}
\begin{document}

\title{Multivariate probabilistic forecasting using Bayesian model averaging and copulas}

\author{Annette M\"{o}ller, Alex Lenkoski\footnote{\noindent \textit{Corresponding author address:} Alex Lenkoski, Institute of Applied Mathematics, Heidelberg University, Im Neuenheimer Feld 294, 69120 Heidelberg, Germany
\newline{E-mail: alex.lenkoski@uni-heidelberg.de}}, and  Thordis L. Thorarinsdottir \\
\textit{Heidelberg University, Germany}}

\maketitle

\begin{abstract}
\noindent
We propose a method for post-processing an ensemble of multivariate forecasts in order to obtain a joint predictive distribution of weather.
Our method utilizes existing univariate post-processing techniques, in this case ensemble Bayesian model averaging (BMA), to obtain estimated marginal distributions.
However, implementing these methods individually offers no information regarding the joint distribution.
To correct this, we propose the use of a Gaussian copula, which offers a simple procedure for recovering the dependence that is lost in the estimation of the ensemble BMA marginals.
Our method is applied to 48-h forecasts of a set of five weather quantities
using the 8-member University of Washington
mesoscale ensemble. We show that our method recovers many well-understood dependencies between weather quantities and subsequently improves calibration and sharpness over both the raw ensemble and a method which does not incorporate joint distributional information.
\end{abstract}

\section{Introduction}

Mesoscale weather forecasting is usually conducted via a forecast ensemble where the ensemble members differ by the boundary conditions and/or the parameterization of the model physics in the numerical weather prediction (NWP) model \citep{LeutbecherPalmer2008}.  While ensemble systems aim to reflect and quantify sources of uncertainty in the forecast, they tend to be biased, and they are typically underdispersed \citep{HamillColucci1997}.  The predictive performance of an ensemble forecast can thus be substantially improved by applying statistical postprocessing to the model output in which the forecasts are corrected in coherence with recently observed forecast errors \citep{WilksHamill2007}.  It has further been argued that the statistical postprocessing techniques should be probabilistic in nature and return full predictive distributions, see e.g. \citet{GneitingRaftery2005}.  State of the art approaches of this type include ensemble model output statistics or nonhomogeneous Gaussian regression \citep{Gneiting&2005, ThorarinsdottirGneiting2010, ThorarinsdottirJohnson2011} and kernel dressing or ensemble Bayesian model averaging \citep{Raftery&2005, Fortin&2006, Sloughter&2007, SloughterGneitingRaftery2010}.

However, many of these techniques consider only a single weather quantity at a fixed look-ahead time, without taking potential spatial dependencies into account.  The postprocessed forecasts may thus violate the multivariate correlation structure of the original ensemble forecasts and the observations.  For low-dimensional multivariate settings, the correlation structure can be modelled directly with a parametric model.  Parametric postprocessing techniques that model the spatial dependencies explicitely through a geostatistical model have e.g. been proposed for temperature \citep{Berrocal&2007} and precipitation \citep{Berrocal&2008}.  Similarly, \citet{Pinson2011}, \citet{Sloughter&2011}, and \citet{Schuhen&2012} consider bivariate probabilistic forecasts for wind vectors.

In higher dimensions, joint parametric modelling becomes cumbersome, especially when the marginal distributions are assumed to be of different types as is e.g. the case for temperature, precipitation, and wind speed.  For such data, copula methods are advantageous as they allow for independent modelling of the marginal distributions and the multivariate dependence structure of the rank statistics, see e.g. \citet{GenestFavre2007}.  We propose a multivariate postprocessing framework where in a first step, established Bayesian model averaging (BMA) methodology is applied independently to each weather variable to obtain calibrated and sharp marginal predictive distributions.  In a second step, the marginal distributions are conjoined in a multivariate framework using a Gaussian copula model, see e.g. \citet{Hoff2007}.

We apply our method to $48$-hour ahead forecasts of five weather variables obtained from the eight-member University of Washington mesoscale ensemble \citep{EckelMass2005} at $60$ observation locations in the North American Pacific Northwest in 2008.  The variables we consider are daily maximum and minimum temperature, sea level pressure, precipitation accumulation, and maximum wind speed.  The marginal predictive distributions of temperature and pressure are defined on the entire real axis while wind speed takes values on the positive real axis only.  Furthermore, the marginal predictive distribution for precipitation accumulation takes values on the non-negative real axis with an additional point mass in zero.  While estimation of the marginal distribution requires incorporation of these features, the Gaussian copula remains largely agnostic to such factors, highlighting its flexibility when working distributions with mixed marginals \citep{Hoff2007}.  An example of the resulting multivariate predictive distribution is given in Figure~\ref{fig:joint}.

Copula methods are widely used for prediction problems in hydrology, see e.g. \citet{GenestFavre2007}, \citet{SchoelzelFriederichs2008}, \citet{KaoGovindaraju2010} and references therein.  In the context of multivariate statistical postprocessing, two discrete copula approaches have been proposed where the multivariate rank structure is inherited from either past observations or the ensemble forecasts.  The Schaake Shuffle \citep{Clark&2004} reorders a postprocessed ensemble based on past observations in order to recover the multivariate rank structure in the observed data.  Ensemble copula coupling (ECC) \citep{Schefzik2011}, on the other hand, constructs multivariate samples from the marginal predictive distributions in such a way that the multivariate rank structure of the original NWP ensemble is preserved.  Our work, by contrast, uses a Gaussian model for dependence, which requires estimation of a single model component (the multivariate correlation matrix).  We feel this approach has the potential to scale considerably better than approaches based on estimation of multivariate ranks, which would require a considerable number of observations in high dimensions to estimate properly. Note that the method proposed by \cite{Pinson2011} may also be considered a copula approach.

The paper is organized as follows.  Description of the data and the methods are given in Section~\ref{sec:methods}.  We shortly review the univariate BMA approaches and the multivariate verification methods that we apply and give a detailed description of the Gaussian copula approach.  The results of the case study are presented in Section~\ref{sec:results} and the paper ends with conclusions in Section~\ref{sec:conclusions}.

\section{Data and methods} \label{sec:methods}

\subsection{Data}\label{sec:data}

In our case study, we employ daily 48-h forecasts based on the University of Washington mesoscale ensemble (UWME; Eckel and Mass 2005) \nocite{EckelMass2005} with valid dates in the calendar year 2008.  The UWME is an eight-member multianalysis ensemble which then was based on the Fifth-Generation Penn State/NCAR Mesoscale Model (MM5) with initial and lateral boundary conditions obtained from operational centers around the world.  Currently, the UWME uses the WRF mesoscale model.  Further information as well as real time forecasts and observations can be found on the website \url{http://www.atmos.washington.edu/~ens/uwme.cgi}.

The forecasts are made on a 12 km grid over the Pacific Northwest region of Western North America.  To obtain a forecast at a given observation location, the forecasts at the four surrounding grid points are bilineraly interpolated to that location.  We consider observation locations in the US states of Washington, Oregon, Idaho, California, and Nevada, see Figure~\ref{fig:corr}.  The daily observations are provided by weather observation stations in the Automated Surface Observing Network \citep{NWS1998}.  We consider 2-m maximum and minimum temperature, sea level pressure, 10-m maximum wind speed, and 24-h precipitation accumulation.  Forecasts and observations are initialized at 00 UTC which is 5pm local time when daylight saving time operates and 4pm local time otherwise.  Quality control procedures as described by \citet{Baars2005} were applied to the entire data set, removing dates and locations with any missing forecasts or observations.

For the calendar year 2008, we consider 60 distinct observation locations which have between $95$ and $271$ days in which all ensemble forecasts and verifying observations were available.  Additional data from 2006 and 2007 were used to provide an appropriate rolling training period for all days in 2008 and to estimate the multivariate correlation structure.

\subsection{Ensemble Bayesian model averaging}\label{sec:BMA}

Bayesian model averaging (BMA) was originally developed as a method to combine predictions and inferences from multiple statistical models \citep{Leamer1978}.  \cite{Raftery&2005} extended the use of BMA to statistical postprocessing for forecast ensembles.  In this context, the method is a kernel dressing approach where each ensemble member $x_k$ is associated with a kernel density $g_k(y|x_k)$.  The ensemble BMA predictive density is then given by a mixture of the individual kernel densities,
\begin{equation}\label{eq:BMA}
f(y|x_1,\ldots,x_K) = \sum_{k=1}^K \omega_k g_k(y|x_k),
\end{equation}
where the weights $\omega_k$ are assumed to be non-negative with $\sum_{k=1}^K \omega_k = 1$.  The choice of the kernel $g_k$ depends heavily on the weather variable of interest:  \cite{Raftery&2005} consider temperature and pressure for which Gaussian kernels seem appropriate, while \cite{SloughterGneitingRaftery2010} apply gamma kernels to wind speed forecasts.

Precipitation has to be treated slightly differently as precipitation observations are non-negative with a large number of zero observations.  \cite{Sloughter&2007} propose a solution to this where the kernel density $g_k$ is modelled in two parts,
\[
g_k(y|x_k) =  \pz(y=0|x_k)\mathbbm{1}\{y=0\} +  \pz(y > 0|x_k) h_k(y^{1/3}|x_k)\mathbbm{1}\{y > 0\},
\]
where $h_k$ is a gamma density and
\[
\pz(y=0|f_k) = \frac{\exp(a_{0k}+a_{1k}x_k^{1/3}+a_{2k} \delta_k)}{1 + \exp(a_{0k}+a_{1k}x_k^{1/3}+a_{2k} \delta_k)}
\]
with $\delta_k=1$ if $x_k=0$ and $\delta_k=0$ otherwise.  Note that $h_k$ is a predictive density for the cube root of the precipitation amount.  However, the resulting probabilistic forecast can easily be expressed in terms of the original amounts.

Table~\ref{tab:BMA models} gives an overview over the different BMA models we consider, as well as the associated link functions for the mean value and the variance of each kernel density.  Further variants of the ensemble BMA method which will not be considered here include the work by \cite{RoquelaureBergot2008}, \cite{Bao&2010}, and \cite{ChmieleckiRaftery2010}.  For the parameter estimation, we apply the {\tt R} package {\tt ensembleBMA} which provides estimation methods for all the models listed in Table~\ref{tab:BMA models} \citep{R2011,Fraley&2011}.

\begin{table}
\centering
\caption{The ensemble BMA kernel functions for the different weather variables and the associated link functions for each mean value and variance.  The gamma distribution is parameterized in terms of shape and scale with mean $\alpha \beta$ and variance $\alpha \beta^2$.}
\label{tab:BMA models}
\smallskip
\begin{tabular}{lllll}
\toprule
Variable & Range & Kernel & Mean & Variance \\
\midrule
Temperature & $y \in \rz$ & $\mathcal{N}(\mu_k,\sigma_k^2)$ & $b_{0k} + b_{1k} x_k$ & $\sigma^2$ \\
Pressure & $y \in \rz$ & $\mathcal{N}(\mu_k,\sigma_k^2)$ & $b_{0k} + b_{1k} x_k$ & $\sigma^2$ \\
Wind speed & $y \in \rz_{+}$ & $\Gamma(\alpha_k,\beta_k)$ & $b_{0k} + b_{1k} x_k$ & $c_0 + c_1 x_k$ \\
Precipitation amount & $y^{1/3} \in \rz_+$ & $\Gamma(\alpha_k,\beta_k)$ & $b_{0k}+b_{1k}x_k^{1/3}$ & $c_0 + c_1 x_k$\\
\bottomrule
\end{tabular}
\end{table}

\subsection{Gaussian Copulas} \label{sec:gcop}
The ensemble BMA methods discussed in Section~\ref{sec:BMA} have proven to work well for post-processing ensemble forecasts of univariate quantities.  However, by using these methods for each quantity individually, no attention has been paid to the joint distribution of weather quantities.  In this section, we outline a Gaussian copula approach that allows us to recover the dependence between weather quantities and construct a post-processed joint distribution.\\
\indent Suppose that we have $p$ weather quantities of interest, with marginal distributions $F_1,\dots,F_p$, where
\begin{equation}\label{eq:marginalcdf}
F_j(y) = \int_{-\infty}^{y} f_j(u|x_{1j},\dots,x_{Kj})du.
\end{equation}
In Equation~\ref{eq:marginalcdf}, $f_j(u|x_{1j}, \dots, x_{Kj})$ represents the ensemble BMA density discussed in Equation~\ref{eq:BMA} for variable $j$, evaluated at $u$ and depending on the ensemble members $x_{1j},\dots,x_{Kj}$.  Now let $\bC$ be a $p\times p$ correlation matrix, i.e. a positive definite matrix with unit diagonal. Under a Gaussian copula, the joint distribution $\bF$ of the weather quantities takes the following form
\begin{equation}\label{eq:GaussianCop}
\bF(y_1, \dots, y_p|\bC) = \Phi_p(\Phi^{-1}(F_1(y_1)),\dots, \Phi^{-1}(F_p(y_p))|\bC),
\end{equation}
where $\Phi^{-1}(\cdot)$ is the inverse cdf of a standard Gaussian distribution and $\Phi_p(\cdot|\Sigma)$ is the cdf of a $p-$variate Gaussian distribution with covariance matrix $\Sigma$.
The Gaussian copula is a particularly tractable type of copula model, as it requires only the marginal distributions $F_1,\dots,F_p$ and the correlation matrix $\bC$ to be fully defined (see \citet{Nelsen2006} for a more detailed account of general copula models).\\
\indent The Gaussian copula lends itself to a useful construction. Let
$$
\mathbf{Z}\sim \mathcal{N}_p(0,\bC)
$$
and for $j = 1, \dots, p$ set
$$
Y_j = F^{-1}_j(\Phi(Z_j)),
$$
where
$$
F_j^{-1}(u) = \sup\{y:F_j(y) \leq u\}
$$
denotes the psuedo-inverse of the marginal $F_j$.  Then we have that $\mathbf{Y} = (Y_1,\dots,Y_p) \sim \bF$.  The construction also highlights that each $Y_j$ is marginally distributed according to $F_j$.\\
\indent The construction above thus creates a link between an observation $\bY$ sampled from $\bF$ and a latent Gaussian factor $\bZ$.  In particular if $F_j$ is a continuous marginal distribution we can immediately see that
$$
Z_j = \Phi^{-1}(F_j(Y_j)).
$$
This indicates that for the majority of weather quantities, which have fully continuous distributions, given $F_j$ and an observed $y_j$ we can directly infer a latent $z_j$. In the case of precipitation, the situation is slightly more nuanced.  In general, suppose that $Y_j \in [0,+\infty)$ where $F_j(0) = \alpha$ with $0 < \alpha \leq 1$ and $F_j$ is otherwise continuously increasing on $(0,+\infty)$.  Then we have that
$$
-\infty < Z_j \leq \Phi^{-1}(\alpha)
$$
when $Y_j = 0$ and
$$
Z_j = \Phi^{-1}(F_j(Y_j))
$$
when $Y_j > 0$.  If we therefore collect several observations $\by^{(1)},\dots,\by^{(T)}$ we may infer latent Gaussian observations $\bz^{(1)},\dots,\bz^{(T)}$ and thereby estimate the matrix $\bC$.\\
\indent Our process for forming the ensemble post-processed joint distribution of weather quantities builds on the logic above.  Suppose that we have a collection $\by^{(1)},\dots,\by^{(T)}$ of observations over $T$ days.  We assume that each
$$
\by^{(t)} \sim \bF^{(t)}(\cdot|\bC)
$$
where
$$
\bF^{(t)}\left(y^{(t)}_{1},\dots, y^{(t)}_p|\bC\right) = \Phi_p\left(\Phi^{-1}\left(F^{(t)}_1(y^{(t)}_1)\right),\dots,\Phi^{-1}\left(F^{(t)}_p(y^{(t)}_p)\right)|\bC\right)
$$
and $F^{(t)}_j$ denotes the ensemble BMA marginal distribution for weather quantity $j$ at timepoint $t$ using the $K$ ensemble members $\bx^{(t)}_1,\dots \bx^{(t)}_K$ at timepoint $t$.  This framework associates each observation $\by^{(t)}$ with its own Gaussian copula $\bF^{(t)}$, but all copulas share one correlation matrix $\bC$.  The assumption of a common correlation matrix $\bC$ can be relaxed, as we outline in the Section~\ref{sec:conclusions}.\\
\indent Using $\by^{(t)}$ and the marginal distributions $F^{(t)}_1,\dots,F^{(t)}_p$ we may then infer a latent Gaussian $\bz^{(t)}$ as discussed above.

While each $\by^{(t)}$ was given a separate distribution $\bF^{(t)}$ we note that the associated $\bz^{(t)}$ are all distributed $\mathcal{N}_p(0,\bC)$.  We therefore use these latent Gaussian observations to estimate $\hat{\bC}$, for instance by taking the sample correlation matrix.  \citet{Hoff2007} discusses more involved methods for estimating $\bC$, we discuss the inclusion of these methods in Section~\ref{sec:conclusions}.\\
\indent Now consider forming a predictive distribution for the timepoint $s$ (coming sometime after $T$), based on the $K$ ensemble members $\bx_1^{(s)},\dots,\bx_K^{(s)}$ and the estimate $\hat{\bC}$.  Our method proceeds by first forming the ensemble BMA predictive marginals $F_1^{(s)},\dots,F_p^{(s)}$ and then setting
$$
\bY^{(s)} \sim \bF^{(s)}(\cdot|\hat{\bC}).
$$
While this joint predictive distribution may not have an easy analytic structure, a sample $\hat{\bY}$ may be obtained by first sampling
$$
\bZ \sim \mathcal{N}_p(0,\hat{\bC})
$$
and then setting
$$
\hat{Y}_j = (F^{(s)}_j)^{-1}(\Phi(Z_j)).
$$
By sampling a large number of $\hat{\bY}$ in this manner, we are able to effectively describe the entire joint predictive distribution.  As noted above, the marginal distributions for each individual quantity of this sample remain the ensemble BMA marginals $F^{(s)}_1,\dots, F^{(s)}_p$.

\subsection{Multivariate forecast verification}

To assess the quality of the multivariate forecasts, we apply the methods described in \cite{Gneiting&2008}.  For inspecting calibration, we use the multivariate rank histogram (MRH) which is a direct generalization of the univariate verification rank histogram or Talagrand diagram \citep{Anderson1996, HamillColucci1997, Talagrand&1997}.  The only challenge lies in defining a multivariate rank order, as no natural ordering exists for multivariate vectors. Here, we use the multivariate ordering described in \cite{Gneiting&2008}.

A forecast is said to be calibrated if the resulting MRH is close to being uniform.  To quantify the deviation from uniformity, we use the discrepancy or reliability index $\Delta$,
\[
\Delta = \sum_{j=1}^{m+1} \Big| \zeta_j - \frac{1}{m+1} \Big|,
\]
where $\zeta_j$ is the observed relative frequency of rank $j$ \citep{DelleMonache&2006}.

The sharpness of a univariate predictive distribution or an ensemble forecast can easily be assessed by the corresponding standard deviation.  In the multivariate case, we employ a generalization of the standard deviation, the determinant sharpness (DS),
\[
\textup{DS } = (\textup{det } \Sigma)^{1/(2d)},
\]
where $\Sigma$ is the covariance matrix of an ensemble or a multivariate predictive distribution for a $d$-dimensional quantity.  For ensemble forecasts, the matrix is generated by empirical variances and correlation of the ensemble.

Scoring rules for the verification of deterministic or probabilistic forecasts are well known and have been widely used in forecast assessment.  We consider multivariate extensions of the absolute error and the continuous ranked probability score \citep{MathesonWinkler1976, Hersbach2000}.  The absolute error generalizes to the Euclidean error (EE),
\[
\textup{EE}(F,\mathbf{y}) = \|\boldsymbol\mu-\mathbf{y}\|,
\]
where $\boldsymbol\mu$ is the median of $F$.  For an ensemble or a sample from a continuous distribution, $\boldsymbol\mu$ is defined as the vector that minimizes the sum of the Euclidean distance to the individual forecast vectors,
\[
\min_{\boldsymbol\mu} \Big\{ \sum_{i=1}^m \| \boldsymbol\mu - \mathbf{x}_i \| \Big\}.
\]
The vector $\boldsymbol\mu$ can be determined numerically using the algorithm described in \cite{VardiZhang2000} as implemented in the {\tt R} package {\tt ICSNP}

For a generalization of the continuous ranked probability score, \cite{GneitingRaftery2007} introduce the energy score (ES),
\[
\textup{ES}(F,\mathbf{y})= E_F || \mathbf{X} - \mathbf{y} || - \frac{1}{2} \; E_F || \mathbf{X} - \mathbf{X}' ||,
\]
where $||\cdot ||$ denotes the Euclidean norm and $\mathbf{X}$ and $\mathbf{X}'$ are independent random vectors with distribution $F$.  If $F$ is the cumulative distribution function associated with a forecast ensemble of size $m$, the energy score can be computed as
\[
\textup{ES}(F, \mathbf{y}) = \frac{1}{m} \; \sum_{j=1}^m \; || \mathbf{x}_j - \mathbf{y} || -
\frac{1}{2 m^2} \; \sum_{i=1}^m \, \sum_{j=1}^m \; ||\mathbf{x}_i - \mathbf{x}_j ||.
\]
Generally, the energy score may be approximated by
\[
\textup{ES}(F, \mathbf{y}) \approx \frac{1}{n} \; \sum_{j=1}^n \; || \mathbf{x}_j - \mathbf{y} || -
\frac{1}{2 n} \; \sum_{j=1}^n \; ||\mathbf{x}_j - \mathbf{x}'_j ||,
\]
where $\{\mathbf{x}_j\}_{j=1}^n$ and $\{\mathbf{x}'_j\}_{j=1}^n$ are two independent samples from $F$.

We assign the forecasting methods a score by averaging the scoring rules over all locations and time points in the test set.  Both the energy score and the Euclidean error are negatively oriented such that a smaller score indicates a better predictive performance.  The values of the weather variables we consider are given on scales which vary by several orders of magnitude.  For this reason, we normalize the components before we calculate the scores using the observed mean values and standard deviations over the test set.

\section{Results} \label{sec:results}

We now give the results of applying the established BMA framework together with our multivariate Gaussian copula approach to maximum and minimum temperature, sea level pressure, maximum wind speed, and precipitation over the North American Pacific Northwest in 2008.  See Section~\ref{sec:data} for a detailed description of the data.  The BMA univariate postprocessing is applied at each observation location separately.  Based on an exploratory analysis using a subset of the data set, we use a 40-day sliding training period for the parameter estimation.  That is, the training period consists of the 40 most recent days prior to the forecast for which ensemble output and verifying observations were available.  In terms of calendar days, this period typically corresponds to more than 40 days.

\subsection{Results at Sea-Tac Airport}

To show the behavior of our methodology in-depth, we first focus on the KSEA observation station, located at Sea-Tac Airport, a major transportation hub in the area.  Using all available data from 2007, we run the ensembleBMA methodology for each of the five variables as described in Section~\ref{sec:BMA}.  We then use the observations for these data and the estimated marginal distributions to infer a latent vector $\bz$ as described in Section~\ref{sec:gcop}.  This is performed separately for each day in $2007$ and the resulting latent data are then used to estimate a single correlation matrix.  Table~\ref{tbl:cor} shows the entries of this correlation matrix.

\begin{table}
\centering
\caption{Estimated correlation matrix at the station KSEA, Sea-Tac Airport, based on data from the calendar year 2007.}
\label{tbl:cor}
\smallskip
\begin{tabular}{l|d{2.3}d{2.3}d{2.3}d{2.3}d{2.3}}
\toprule
 & \multicolumn{1}{c}{maxwsp} & \multicolumn{1}{c}{precip} & \multicolumn{1}{c}{mintemp} & \multicolumn{1}{c}{maxtemp} & \multicolumn{1}{c}{pressure}\\
\midrule
maxwsp & 1 & -0.016 & 0.032 & 0.139 & -0.123\\
precip & -0.016 & 1 & -0.001 & -0.174 & -0.015\\
mintemp & 0.032 & -0.001 & 1 & 0.239 & -0.110\\
maxtemp & 0.139 & -0.174 & 0.239 & 1 & -0.203\\
pressure & -0.123 & -0.015 & -0.110 & -0.203 & 1\\
\bottomrule
\end{tabular}
\end{table}

The correlations estimated in Table~\ref{tbl:cor} show several clear patterns of interaction.  We see a strong negative correlation between pressure and both maximum and minimum temperatures.  This is in line with the understood inverse relationship between temperature and pressure systems.  We also see an intuitive positive correlation between the minimum and maximum temperatures.

This estimated correlation matrix is then carried forward into an analysis of 2008, which we use for verification.  For each day with observations in 2008, we again run ensembleBMA individually for each weather quantity to obtain estimated marginal predictive distributions.  We then use the estimated correlation matrix and these marginal distributions to obtain 20,000 samples from the joint predictive distribution for each day, as described in Section~\ref{sec:gcop}.

\begin{figure}
\centering
\caption{Estimated joint predictive distribution for January 1st, 2008 at the KSEA observation station along with ensemble predictions (circles) and verifying observation (square).  In the pairwise plots, lighter areas correspond to regions of higher probability mass.  The diagonal shows the marginal predictive distribution for each quantity.  Wind speed is given in meters per second, precipitation in millimeters, temperature in degrees C, and pressure in millibars.}\label{fig:joint}
\includegraphics[width = 5in]{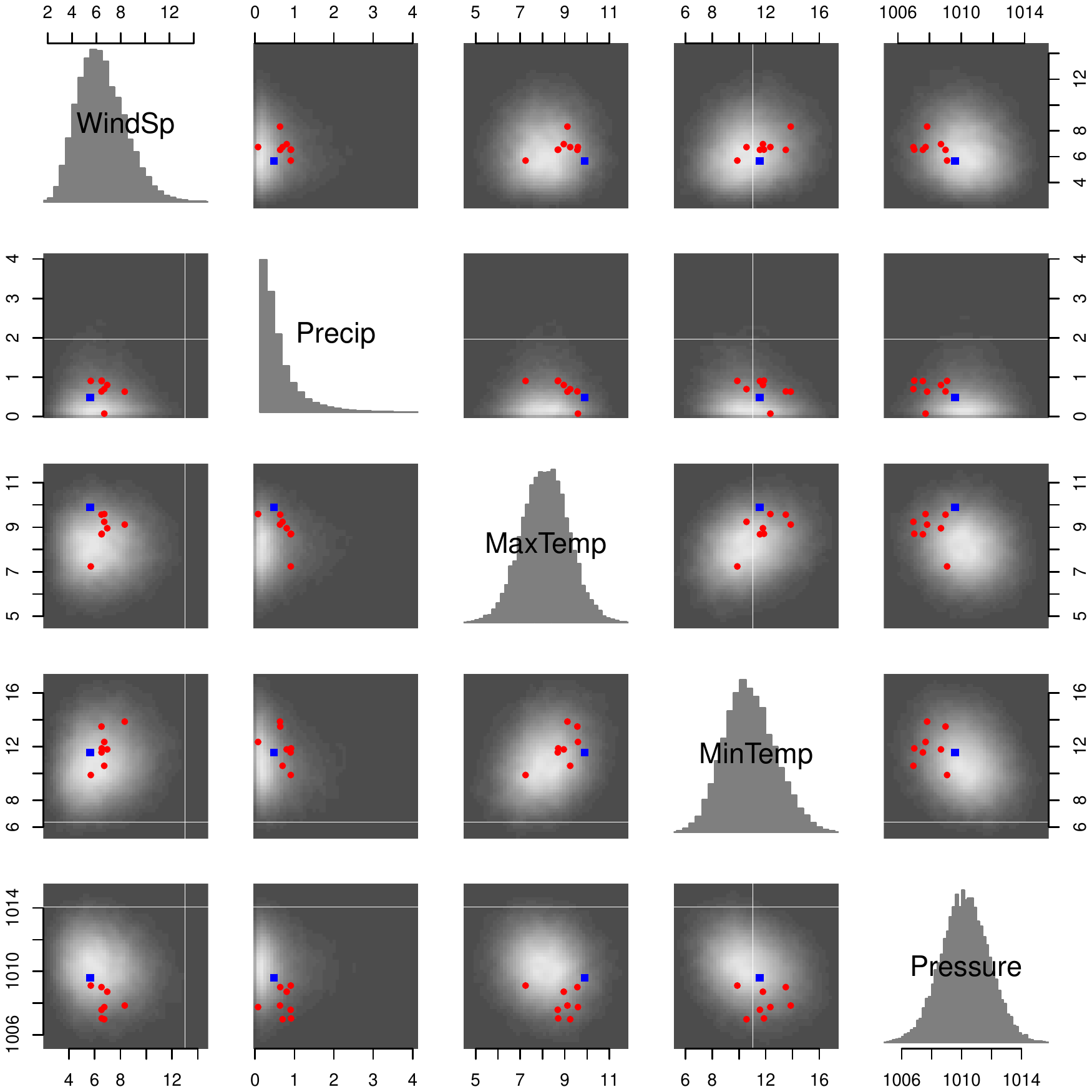}
\end{figure}

Figure~\ref{fig:joint} shows a pairwise plot of this joint predictive distribution, for the date January 1st, 2008.  For each pair of variables, the figure shows a heatmap indicating regions of significant probability mass -- with lighter regions being higher values -- as well as points showing the 8 ensemble members (circles) and the observed level (a square).  The marginal predictive distribution, given by the ensembleBMA methodology, is shown for each variable along the diagonal.

In this figure we can see that the correlation structure in Table~\ref{tbl:cor} has been carried over to the predictive distribution: the positive correlation between maximum and minimum temperatures is evident as well as the negative correlation of each of these quantities with pressure.  Further, we can see the effect of post-processing, as the predictive distributions are often centered away from the ensemble members (indicating the bias correction properties) and display greater spread in the distribution than is evident in the ensemble.

Table~\ref{tbl:scoreksea} shows that according to a number of verification metrics, the Gaussian copula approach yields an improvement in predictive performance.  The table compares the Gaussian copula approach to an alternative where dependence is not modeled, which we call the independence approach, as well as the raw UWME ensemble averaged over 2008 for the KSEA observation station.  Note that the Gaussian copula and independence approaches have the same marginal distributions, and thus differ only in the manner in which the joint distribution is constructed.

\begin{table}
\centering
\caption{Predictive performance of the copula methodology, the independence approach, and the raw UWME ensemble at the KSEA observation station.  Results are averaged over the 271 days in 2008 for which forecasts and verifying observations were available.  The performance is measured by the energy score (ES), the Euclidean error (EE), the reliability index ($\Delta$), and the determinant sharpness (DS).}\label{tbl:scoreksea}
\smallskip
\begin{tabular}{ld{1.3}d{1.3}d{1.3}d{1.3}}
\toprule
 & \multicolumn{1}{c}{ES} & \multicolumn{1}{c}{EE} & \multicolumn{1}{c}{$\Delta$} & \multicolumn{1}{c}{DS}\\
\midrule
UWME         & 0.938 & 1.081 & 0.185 & 0.566\\
Independence & 0.637 & 0.982 & 0.047 & 7.516\\
Copula       & 0.636 & 0.982 & 0.019 & 6.971\\
\bottomrule
\end{tabular}
\end{table}

Both the copula and independence approaches improve considerably on the raw ensemble in all metrics except the determinant sharpness (DS).  This is not surprising since the 8 members of the raw ensemble will impart greater sharpness at the expense of calibration.  We also see that the copula approach improves on the independence approach for all metrics.  The values of the reliability index and the DS show that the predictive distribution for the copula approach is both better calibrated and somewhat sharper than for the independence approach, a consequence of using a non-diagonal correlation matrix.  The Euclidean scores are essentially the same for the two approaches, since they can largely be expected to return similar median values.  The combination of similar median and improved sharpness leads the copula approach to have a lower energy score.

\begin{figure}
\centering
\subfigure[UWME]{\includegraphics[width=1.7in]{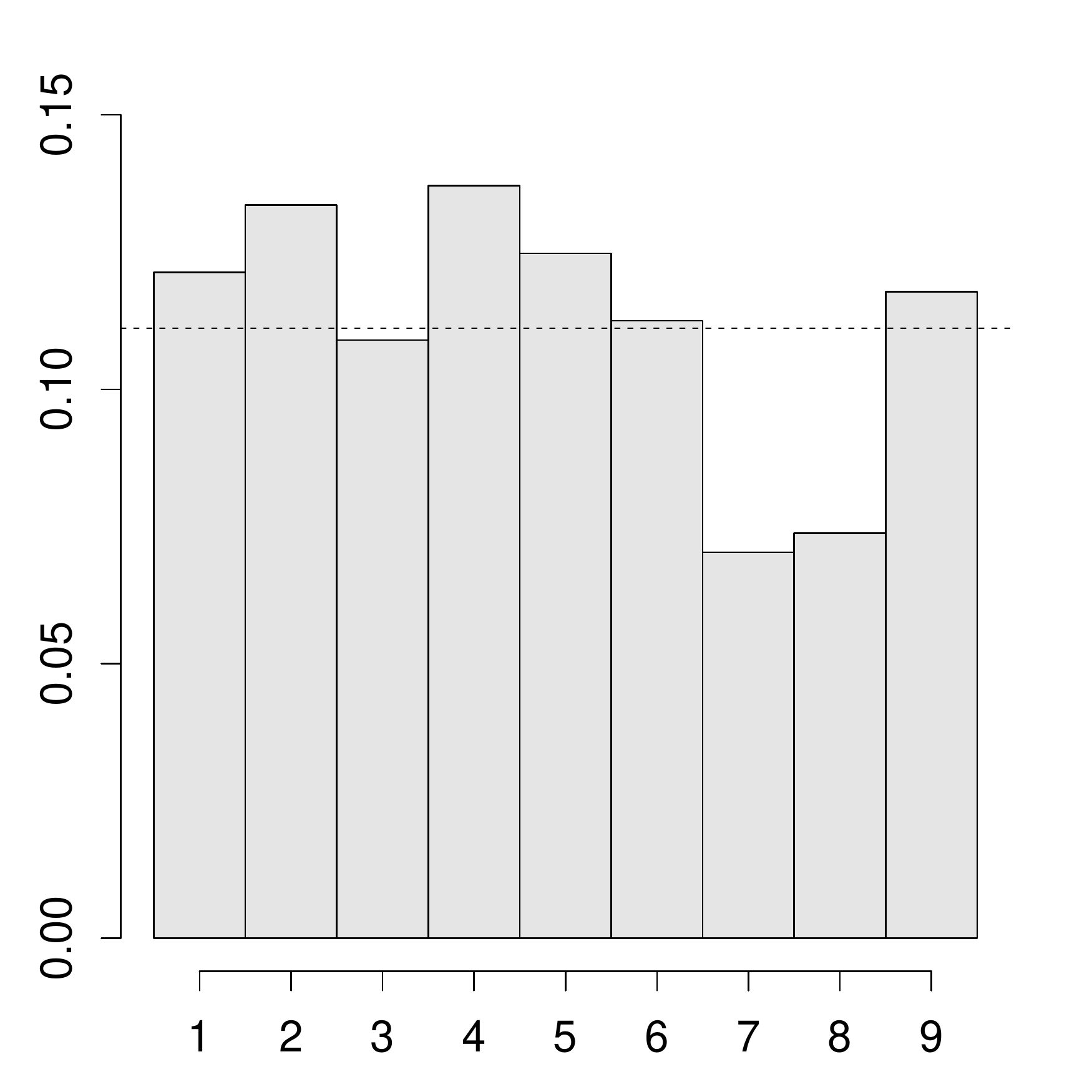}}
\subfigure[Independence]{\includegraphics[width=1.7in]{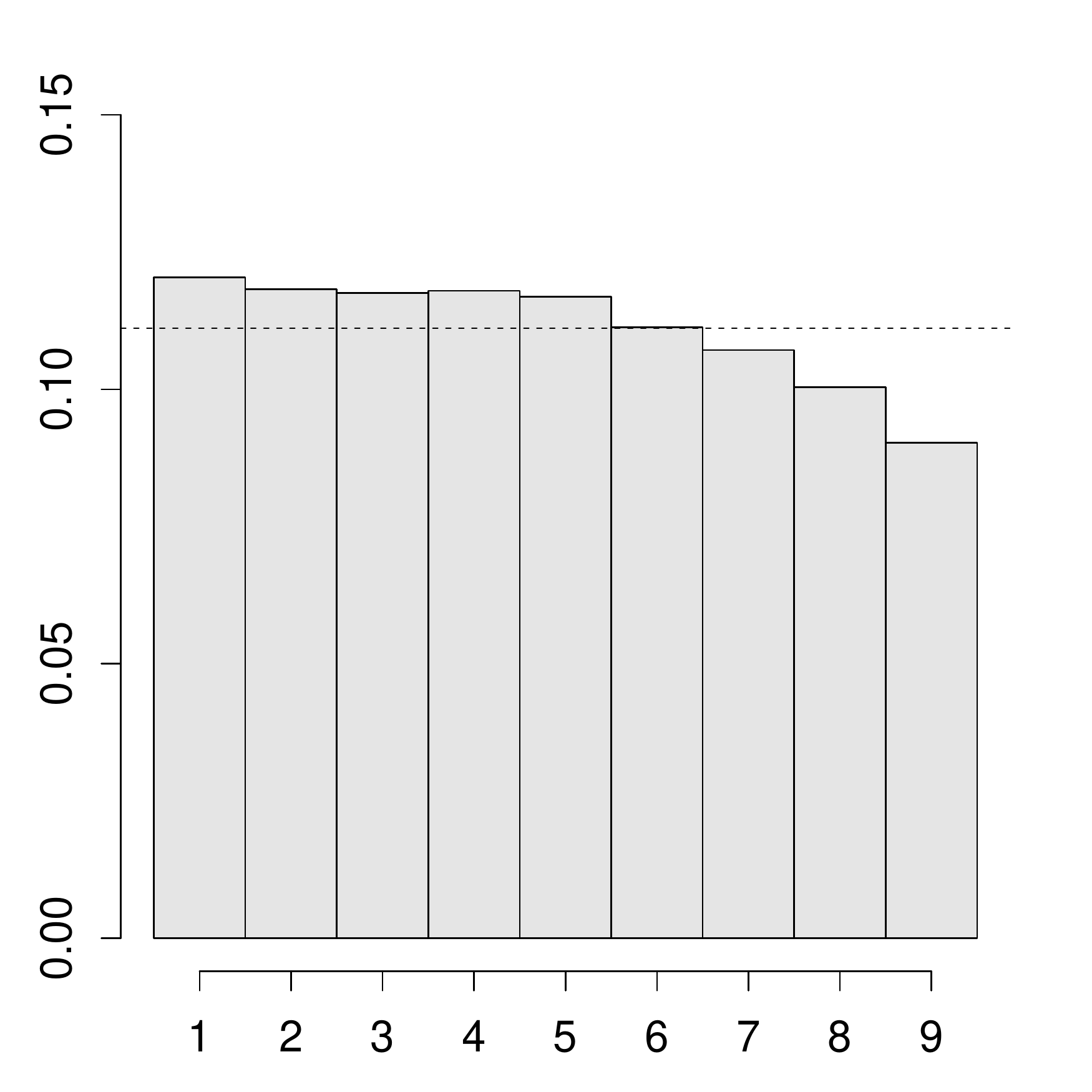}}
\subfigure[Copula]{\includegraphics[width=1.7in]{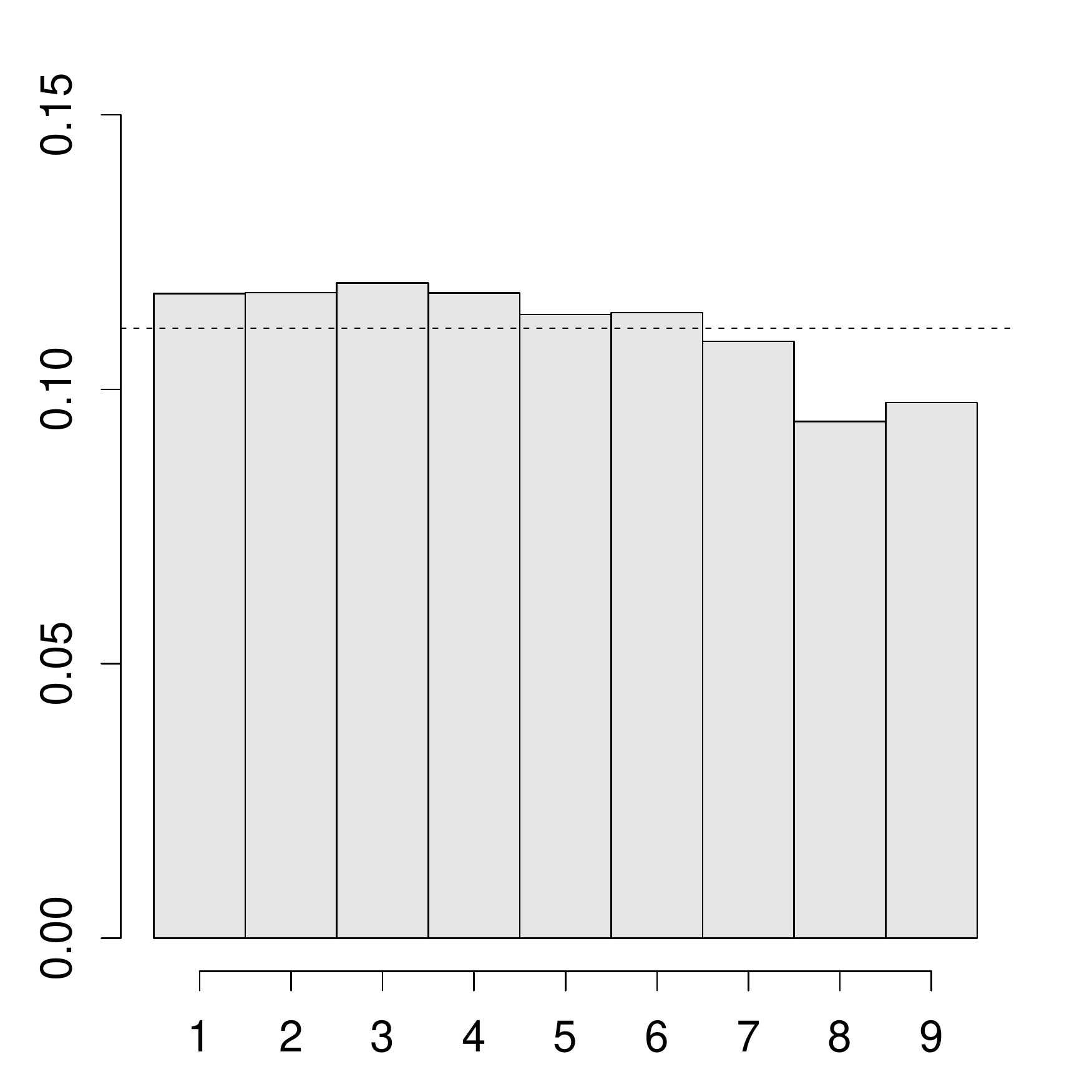}}
\caption{Multivariate rank histograms for the copula and independence approaches as well as the UWME ensemble for the KSEA observation station over the 295 days in 2008 for which forecasts and verifying observations were available.}\label{fig:mvksea}
\end{figure}

Figure~\ref{fig:mvksea} shows the multivariate rank histograms for the two approaches, as well as for the raw ensemble.  The figure shows that both methods improve calibration considerably over the raw ensemble.  However, as shown in the figure, in the independence approach the final bins are somewhat less filled than under the copula approach, though neither returns a completely uniform rank histogram.

\subsection{Results over the Northwest US}

\begin{figure}[!hbpt]
\centering
\subfigure[Minimum and maximum temperature]{\fbox{\includegraphics[width=0.48\textwidth]{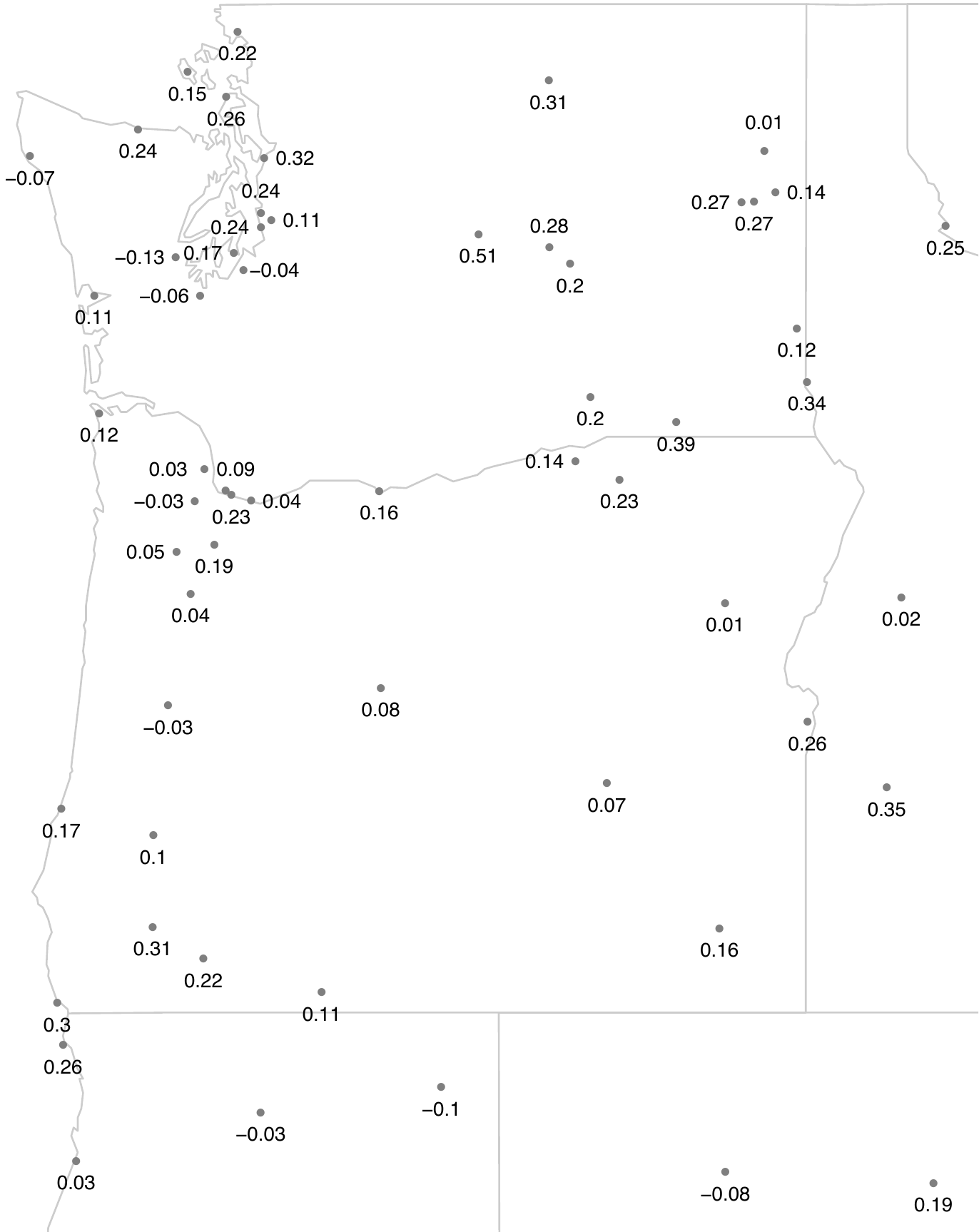}}}
\subfigure[Minimum temperature and pressure]{\fbox{\includegraphics[width=0.48\textwidth]{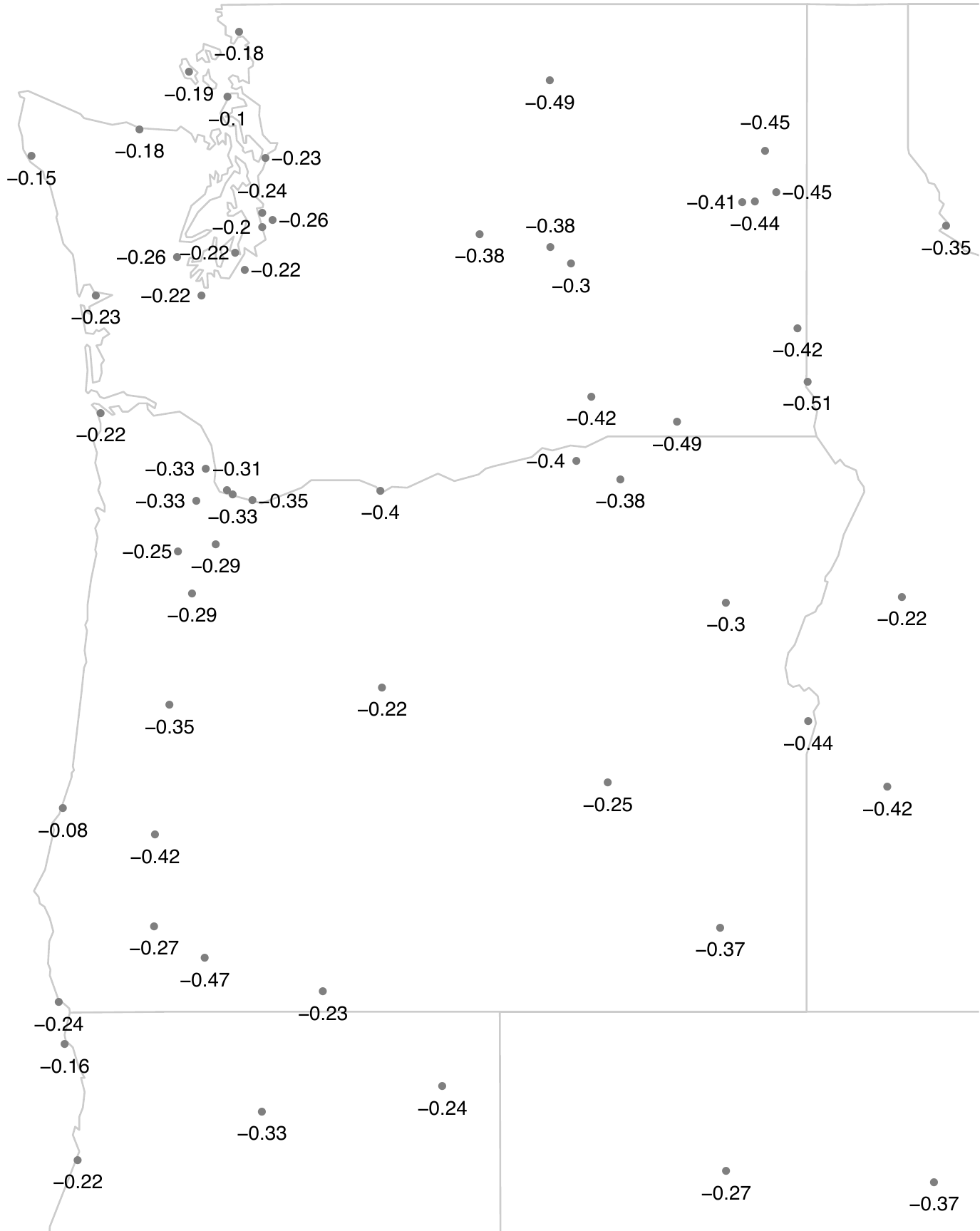}}}
\caption{Estimated correlation between (a) minimum and maximum temperature, and (b) minimum temperature and pressure at $60$ observation stations in the Northwest US using 2007 data and the Gaussian copula approach.}\label{fig:corr}
\end{figure}

A similar analysis as that above was run for 60 separate observation stations in the Northwest US.  First, ensembleBMA was run individually for each station, day and weather quantity during the period of 2007.  Verifying observations were then used to estimate a correlation matrix separately for each observation station.  While these estimates were performed locally, there is considerable agreement in estimated correlations between individual stations.  Figure~\ref{fig:corr}(a) shows the pairwise estimated correlation between minimum and maximum temperature at each observation station.  This plot shows that the majority of estimates are positive, as expected, with correlations as high as $0.51$.

Figure~\ref{fig:corr}(b) shows a similar plot, but with the correlation between minimum temperature and pressure plotted for each observation station.  While the previous figure showed some similarity in estimated correlations, this figure shows a considerable agreement between observation stations.  We see that all values are quite negative, with the majority of estimates between $-0.2$ and $-0.4$.  Furthermore, there appears to be a spatial component to the estimates.  The values near the top of the Puget Sound are all roughly between $-0.1$ and $-0.2$, while those closer to the Seattle/Tacoma area are grouped between $-0.22$ and $-0.26$.  A tight group of observation stations in the Columbia River Valley, on the border of Washington and Oregon, all have correlations between $-0.31$ and $-0.35$ and finally those in Eastern Washington and Eastern Oregon exhibit stronger correlations typically below $-0.4$.  Figures~\ref{fig:corr}(a) and ~\ref{fig:corr}(b) therefore suggest that an important feature of the joint distribution is captured through the Gaussian copula methodology that is ignored in the independence approach.

\begin{figure}
\centering
\subfigure[UWME]{\includegraphics[width=1.7in]{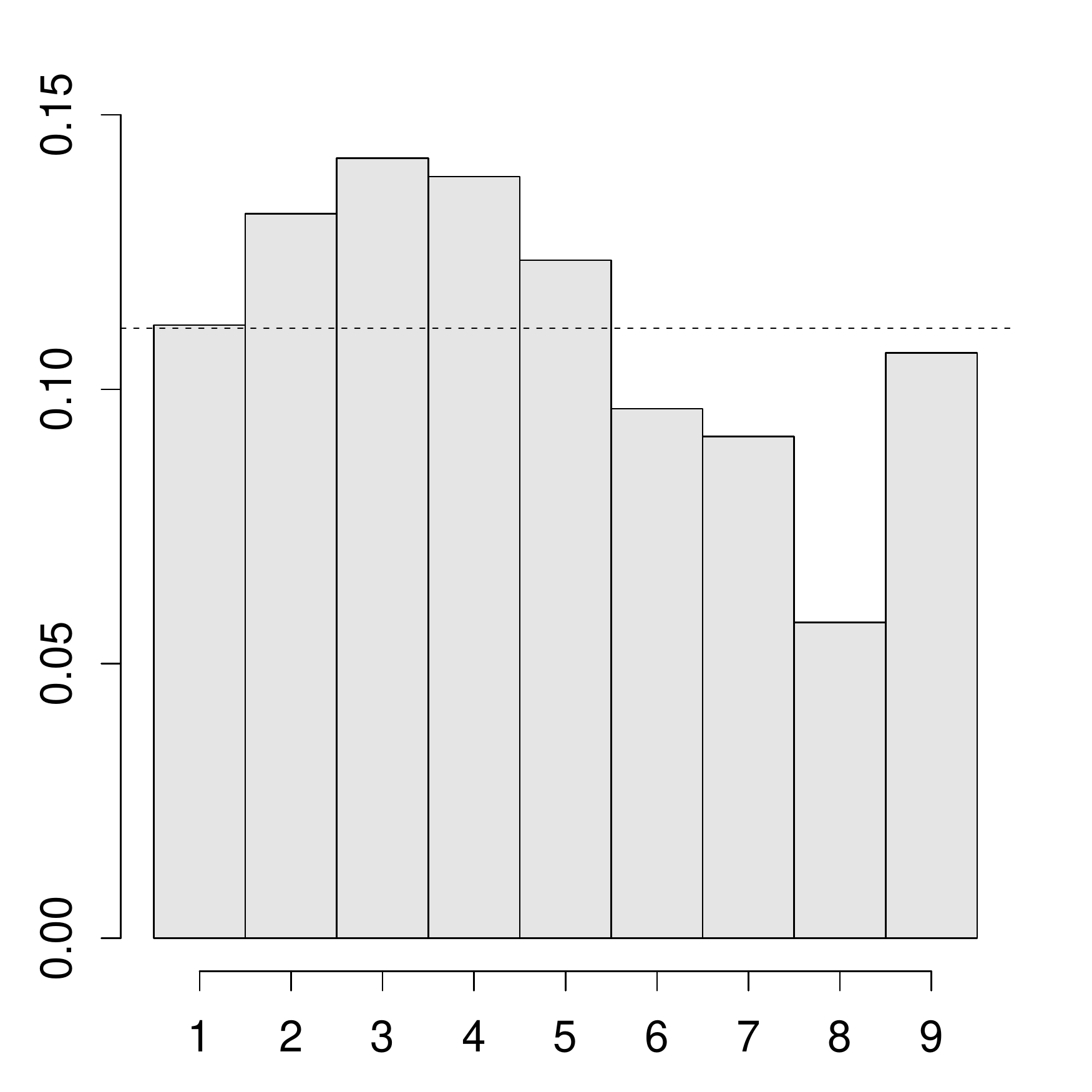}}
\subfigure[Independence]{\includegraphics[width=1.7in]{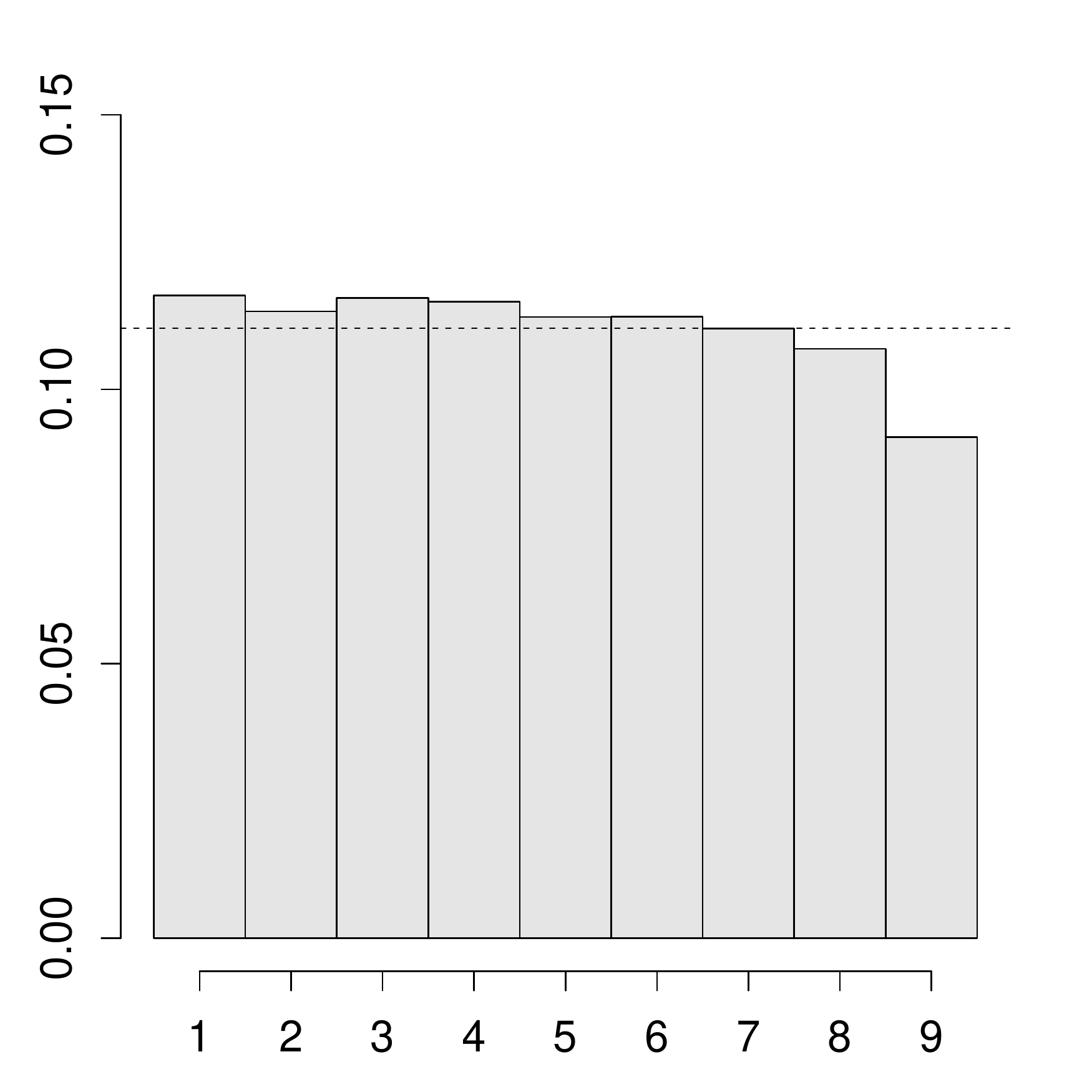}}
\subfigure[Copula]{\includegraphics[width=1.7in]{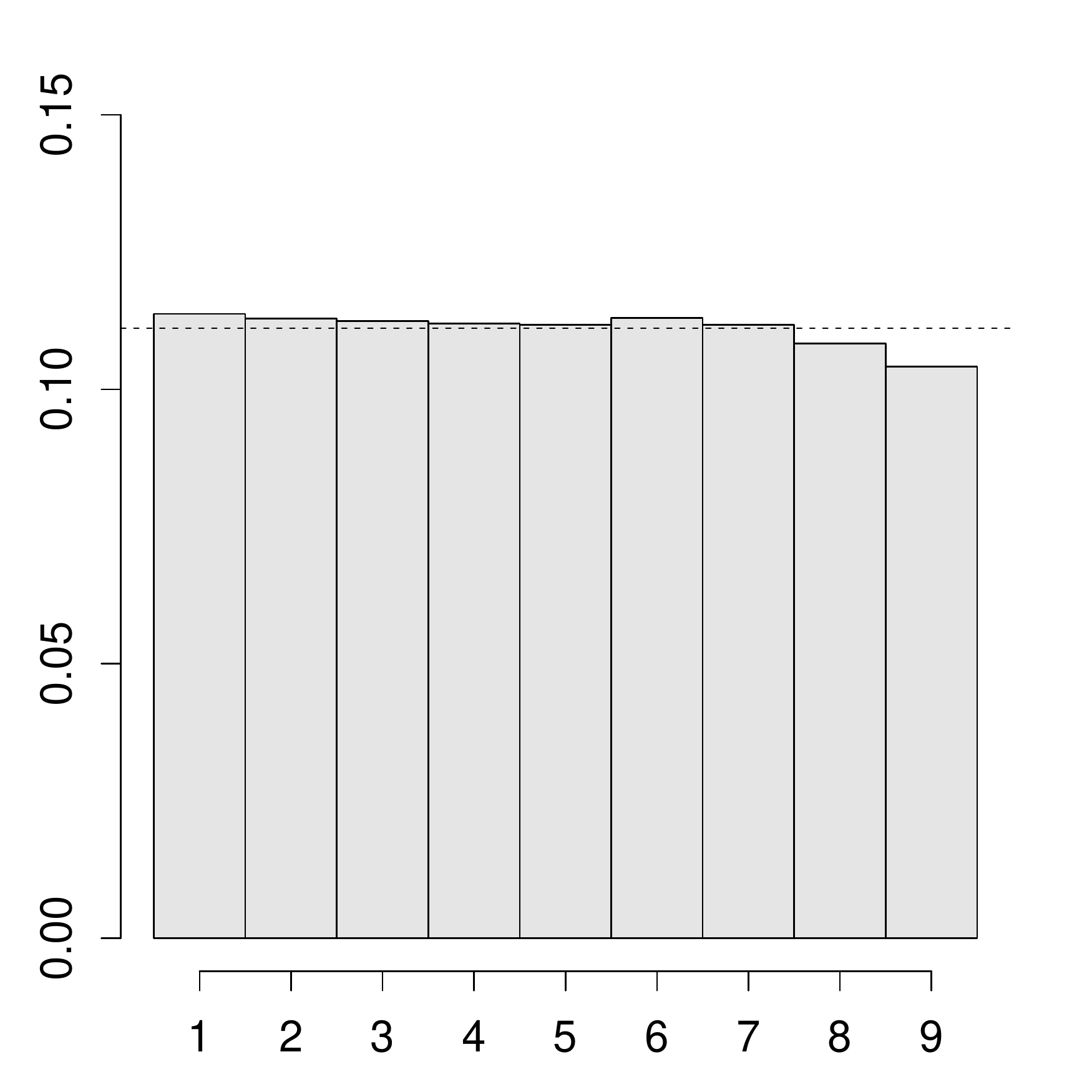}}
\caption{Multivariate Rank Histograms for the copula and independence approaches as well as the raw UWME ensemble taken over all available observations at $60$ observation stations in the Northwest US over the year 2008.}\label{fig:mvall}
\end{figure}

Figure~\ref{fig:mvall} compares the multivariate rank histograms (averaged over all observation stations) for the copula and independence approaches as well as the raw UWME ensemble.  We see a similar result across all stations as for the KSEA station above: namely that the highest bin is under-occupied in the independence approach in comparison to the copula approach.  Furthermore, both methods improve the calibration considerably compared to the raw ensemble.

Table~\ref{tbl:scoreall} presents verification scores averaged over all the $60$ stations and all available days in 2008.  The results here are broadly consistent with those reported for KSEA.  We see that the determinant sharpness is improved in the copula versus independence approach, while the Euclidean score is essentially the same.  These two factors lead to an improvement in the energy score in the copula approach.  Calibration also appears to be improved, as shown through lower energy score and reliability index for the copula approach.

\begin{table}
\centering
\caption{Predictive performance of the copula and independence approaches as well as the raw UWME ensemble.  Results are averaged over $60$ observation stations in the Northwest US and all days in 2008 for which forecasts and verifying observations were available.  The predictive performance is measured by the energy score (ES), the Euclidean error (EE), the reliability index ($\Delta$), and the determinant sharpness (DS).}\label{tbl:scoreall}
\smallskip
\begin{tabular}{ld{1.3}d{1.3}d{1.3}d{1.3}}
\toprule
 & \multicolumn{1}{c}{ES} & \multicolumn{1}{c}{EE} & \multicolumn{1}{c}{$\Delta$} & \multicolumn{1}{c}{DS}\\
\midrule
UWME          & 0.881 & 1.061 & 0.161 & 0.811 \\
Independence & 0.586 & 0.914 & 0.071 & 1.945 \\
Copula       & 0.585 & 0.914 & 0.066 & 1.905 \\
\bottomrule
\end{tabular}
\end{table}

\section{Conclusions} \label{sec:conclusions}
\indent We have proposed a method for constructing a joint predictive distribution based on post-processing an ensemble of forecasts for multiple weather quantities.  Our method utilizes existing techniques for post-processing univariate weather quantities and then leverages a Gaussian copula to tie these individual marginal distributions together.  The method is relatively simple, it requires little additional computational effort after the univariate marginals are formed and it retains the marginal distributions learned from ensemble BMA.  We have then shown that the method yields a calibrated and sharp distribution, using data from the Pacific Northwest.\\
\indent In this paper, we focused on using the ensemble BMA methodology to construct marginal predictive distributions.  This was done since the methodology has already been established for several interesting weather quantities and software exists to implement these methods.  However, in practice any method for forming marginal distributions could replace the ensemble BMA framework and essentially nothing of the overall methodology would be affected.  This highlights the flexibility of the copula approach.\\
\indent Alternative approaches to constructing joint distributions such as the Schaake shuffle and the ECC approach use a discrete copula to construct the joint distribution.  These methods learn a multivariate rank structure by investigating ranks in historical observations (Schaake) or the ensemble (ECC).  We feel our approach offers a useful alternative to these methods for two reasons.  First, the joint distribution is modeled using a Gaussian distribution.  This does not require subsequent samples from the predictive distribution to strictly obey any observed rank structure and, in our opinion, may scale better to high dimensions.  The statistics literature is currently investigating these questions (see \citet{Hoff&2011}) and subsequent work should be undertaken to compare these methods in the ensemble post-processing context.  The second feature of our method is that it aims to model the joint distribution \emph{after} post-processing.  By running the ensemble BMA approach and then learning the latent Gaussian factor using a verifying observation, we are implicitly modeling the joint residual structure implied by the ensemble BMA method itself.  This approach, in our opinion, more appropriately reflects the modeling process that is being performed.\\
\indent Section~\ref{sec:gcop} outlined the basics of Gaussian copula theory and provided a simple algorithm for calculating latent Gaussian factors and the subsequent correlation matrix.  We presented this approach as it is relatively straightforward and captures the main features of interest.  More involved estimation strategies exist, for instance the Bayesian approach of \citet{Hoff2007} and subsequent work should consider these frameworks to assess whether they offer improvements in predictive performance.\\
\indent As noted in Section~\ref{sec:gcop} we assume a constant correlation factor for all observations when estimating the Gaussian copula.  Certainly, such an assumption may be an over-simplification.  More complicated methods that include time-varying correlation factors are possible, but these methods are both more difficult to describe and to estimate.  Further, preliminary investigations on our part revealed essentially no added benefit to considering time-varying correlation models in the example discussed in Section~\ref{sec:results}.  However, it may be useful to reconsider the estimation strategy in the future, especially in higher-dimensional situations.\\
\indent The Gaussian copula has received occasional criticism (see \citet{Mikosch2006}) for not accurately capturing dependence in the tails of multivariate distributions.  This criticism is germane, but not a central concern to the modeling task undertaken in this paper.  Our goal has been to construct a multivariate distribution after post-processing an ensemble of forecasts.  If extreme weather events were possible, it is probable that this information would already be incorporated to some extent in the ensemble itself, and the copula would simply indicate the variability about these extreme values.  Furthermore, the direct modeling task has focused on day-to-day weather prediction, as opposed to focusing on extreme events.  If the joint modeling of extreme weather is the central concern of the forecaster, it is likely that an alternative copula model may be preferable.\\
\indent While we have shown promising results in modeling five weather quantities simultaneously, the long term goal is undoubtedly to model weather jointly in the spatial domain and for multiple variables.  Once spatial factors are added, the dimensionality of the model can increase rapidly and issues related to spatial covariation estimation are necessary to solve.  There have been several recent advances in using fast computational methods for Gaussian Markov random fields \citep{LindRue}, which could prove useful in constructing high-dimensional joint distributions based on post-processed ensemble forecasts.  Our current steps in advancing the methods discussed above have been to merge this literature with that of ensemble post-processing.

\section*{Acknowledgements}
We would like to thank Tilmann Gneiting and Michael Scheuerer for helpful discussions, Jeff Baars for providing the data, and Chris Fraley for invaluable assistance regarding the {\tt ensembleBMA} package.  Annette M\"oller and Alex Lenkoski gratefully acknowledge support by the German Research Foundation (DFG) within the programme ``Spatio-/Temporal Graphical Models and Applications in Image Analysis'', grant GRK 1653.

\bibliography{copula}

\end{document}